\documentstyle[emulateapj]{article}

\begin{document}


\title{Nature and Nurture: a model for soft gamma-ray repeaters}

\author{Bing Zhang$^{1,2,3}$, R. X. Xu$^{4}$, G. J. Qiao$^{5,4}$}

\altaffiltext{1}{Laboratory of High Energy Astrophysics,
 NASA Goddard Space Flight Center, Greenbelt, MD 20771}
\altaffiltext{2}{National Research Council Research Associate}
\altaffiltext{3}{Present address: Department of Astronomy and
 Astrophysics, Pennsylvania State University, University Park, PA
 16803, bzhang@astro.psu.edu}
\altaffiltext{4}{CAS-PKU Beijing Astrophysical Center
 and Astronomy Department, Peking University, Beijing 100871, China}
\altaffiltext{5}{CCAST (World Laboratory) P.O.Box 8730,
 Beijing 100080, China}

\begin{abstract}
During supernova explosions, strange stars with almost bare quark
surfaces may be formed. Under certain conditions, these stars
could be rapidly spun down by the torque exerted by the fossil
disks formed from the fall-back materials. They may also receive
large kicks and hence, have large proper motion velocities. When
these strange stars pass through the spherical ``Oort'' comet
cloud formed during the pre-supernova era, they will capture some
small-scale comet clouds and collide with some comet-like objects
occasionally. These impacts can account for the
repeating bursts as observed from the soft gamma repeaters (SGRs).
According to this picture, it is expected that SGR 1900+14 will
become active again during 2004-2005.
\end{abstract}



\section{Introduction}

Soft Gamma-ray Repeaters (hereafter SGRs) and Anomalous X-ray
Pulsars (hereafter AXPs) are two groups of enigmatic sources. They
share the following properties: 1. They all have long rotation
periods (clustered within 5-12 seconds) and
large spin-down rates (see, e.g. Mereghetti \& Stella
1995\markcite{ms95}; Kouveliotou et al. 1998,
1999\markcite{k98}\markcite{k99}); 2. Most of them are associated
with supernova remnants, indicating that they are young objects
(for reviews, see Hurley 1999\markcite{h99}; Mereghetti
1999\markcite{m99}); 3. No optical, infrared or radio
counterparts have been identified (e.g. Eikenberry \& Dror 
2000\markcite{ed00};
Lorimer \& Xilouris 2000\markcite{lx00}); 4. They all have soft
persistent pulsed X-ray emission with luminosities of $L_{\rm
x} \sim 10^{35}-10^{36}~{\rm ergs~s^{-1}}$, well in excess of the
spin down energy of these sources (e.g. Thompson
2000\markcite{t00} for a review). The main difference between
both types of the objects is that SGRs show occasional soft
gamma-ray bursts while AXPs do not. It is also found that SGRs
usually have larger proper motion velocities than AXPs according
to their relative positions with respect to the cores of their
supernova remnants (Hurley 1999). The main characteristics of the
SGR bursts include: 1. Most of the bursts have super-Eddington
luminosities with $L_{\rm b}\sim 10^{38}-10^{42}~{\rm
erg~s^{-1}}$; 2. The fluence distribution of the bursts 
is a power-law, and there is no correlation between the
burst intensity and the time intervals between the bursts
(G\"og\"us et al. 1999; 2000\markcite{g99}\markcite{g00}); 3. Two
giant flares have been detected from SGR 0526-66 (the March 5,
1979 event) and SGR 1900+14 (the August 27, 1998 event), which
share some common properties (see Thompson 2000 for a review); 4.
Most bursts have soft spectra with characteristic energy around
20-30 keV.

The popular model for SGRs and AXPs is the magnetar model, which
can account for almost all the phenomena listed above (Duncan \&
Thompson 1992\markcite{dt92}; Thompson \& Duncan 1995,
1996\markcite{td95}\markcite{td96}; Thompson 2000). However, the
differences between SGRs and AXPs are not straightforwardly 
interpreted since these objects are not intrinsically
different objects within the magnetar picture. It also remains
unclear how some other issues, e.g., the non-systematic
discrepancy between the characteristic ages derived assuming 
dipolar spindown and the ages of the
associated supernova remnants, no clear positive dependence
between $L_{\rm x}$ and the polar surface field strength $B_{\rm
p}$, etc., can be properly addressed. On the other hand, a
fossil-disk-accretion model for AXPs recently emerges from the
independent studies by Chatterjee et al. (Chatterjee, Hernquist
\& Narayan 2000\markcite{chn00}; Chatterjee \& Hernquist
2000\markcite{ch00}) and Alpar (1999,
2000)\markcite{a99}\markcite{a00}. The neutron stars in such a
scenario have normal magnetic fields as the Crab pulsar. The model
can interpret the AXP phenomenology well, but the bursts from the
SGRs are difficult to interpret. On the observational ground,
Marsden et al. (2000\markcite{mlrh00}) observed that the
SGRs and the AXPs are located in a much denser environment than
the normal pulsars. They hence argue that the peculiar behaviors
of the SGRs and AXPs may be due to their ``nurture'' from the
environment rather than due to their special ``nature'' (i.e.
magnetars) as compared with the normal pulsars. However, no
plausible idea was proposed to connect the ``nurture'' to the
phenomenology of these sources, especially the bursting behavior
of the SGRs.

In this Letter, we attempt to propose a model to understand the
bursting behavior of the SGRs without introducing the magnetar idea.
We propose that the central objects of the SGRs are
``bare'' strange stars with normal magnetic fields ($10^{12}-
10^{13}$ G). We assume that these strange stars are born directly
from supernova explosions from some massive progenitors, and they
have experienced a spindown history as that having been proposed for
the AXPs within the fossil disk model (Chatterjee et al. 2000; Alpar
2000). According to this model, some fallback materials from the
supernova ejecta will form a fossil disk around the strange star.
The SGRs/AXPs are just such strange/neutron stars that have experienced
the ``propeller'' phase ($r_{\rm c} \ll r_{\rm m} < r_{\rm l}$), and
are now in the ``tracking'' phase ($r_{\rm c} \lesssim r_{\rm m}
< r_{\rm l}$) when infall of the materials onto the surface is
possible and the star is X-ray bright. Here $r_{\rm l}$, $r_{\rm m}$
and $r_{\rm c}$ are the light cylinder, the magnetospheric radius,
and the corotating radius, respectively. 
In our picture, AXPs may be still neutron stars. We will attribute
the SGR bursts to their occasionally collisions with some comet-like 
objects in the dense environment of the SGRs. We will show how 
various SGR properties as reviewed above could be accounted for 
within this picture. Our model differs from some other strange star 
SGR models (e.g. Alcock et al. 1986b; Cheng \& Dai 1998; Dar \& de 
Rujula 2000).

\section{The model}

Strange stars (Haensel, Zdunik \& Schaeffer 1986\markcite{hzs86};
Alcock, Farhi \& Olinto 1986a\markcite{afo86a}) are hypothetical
objects based upon the assumption that strange quark matter is
more stable than nuclear matter (Witten 1984\markcite{w84}; Farhi
\& Jaffe 1984). Though the existence of such stars are still
subject to debate, some evidence in favor of strange stars has
recently been collected (e.g. Li et al. 1999a, 1999b; Titarchuk \&
Osherovich 2000). Strange stars can be either bare or have normal
matter crusts (Alcock et al. 1986a). They can be formed
directly during or shortly after some supernova explosions when
the central density of the proto-neutron stars is high enough to
induce phase conversion (e.g. Dai, Peng \& Lu 1995; Xu, Zhang \&
Qiao 2000). If a strange star is born directly from a supernova
explosion, it is likely that the star might be almost bare (Xu et
al. 2000). Some radio pulsars may be such strange stars with
exposed bare quark surfaces (Xu, Qiao \& Zhang 1999).

There are three main motivations for us to choose (bare) strange
stars rather than neutron stars to interpret the SGRs. 1. A
prominent feature of the SGR bursts is their
super-Eddington luminosities. This feature has been regarded as a
strong support to the magnetar model, since superstrong magnetic
fields may considerably suppress the Thompson cross section and
consequently raise the Eddington limit to several orders of
magnitude higher (Paczynski 1992; Thompson \& Duncan 1995).
However, the luminosities of the most luminous events, e.g., the
initial spike of the March 5 event with $L\sim 10^{44}{\rm ergs~s
^{-1}}$, are still above the enhanced Eddington limit. An
important merit of bare strange stars is that they are not
subject to Eddington limit at all since the bulk of the star
(including the surface) is bound via strong interaction rather
than gravity (Alcock et al. 1986a). This presents a clean
interpretation to the super-Eddington luminosities of the SGRs,
as long as the impacts are not in the polar cap region where the
accretion flow from the fossil accretion disk is channeled. 2. As
criticized by Thompson \& Duncan (1995), the impacting model for
neutron stars suffers the baryon contamination problem. The
impact may load too much baryonic matter to cause adiabatic
dilution of photons in an expanding fireball to energies well
below the hard X-ray and $\gamma$-ray band. A bare strange star
can naturally evade such a criticism, since the infall matter
will be essentially converted into strange quark matter within a
very short period of time ($\sim 10^{-7}$ s, Dai et al. 1995) 
when they penetrate into the star. A new-born bare strange star 
may have a very thin normal matter atmosphere (Xu et al. 2000), 
which is far less than the amount required to pollute the 
fireball. 3. Observationally, SGRs tend
to have larger proper motion velocities ($\sim 1000 {\rm
km~s^{-1}}$) than normal pulsars and AXPs. Though we do not
attempt to propose a detailed ``kick'' theory in the present
Letter, we note that the formation of a strange star rather than
a neutron star may potentially pose some possibilities to
interpret the large proper motion velocities of SGRs. Present
kick theories invoke either hydrodynamically-driven or
neutrino-driven mechanisms (Lai 2000). For the former, the kick 
arises from presupernova g-mode perturbations amplified during 
the core collapse, leading to asymmetric explosion (Lai \&
Goldreich 2000). We note that the formation of a
strange star is a two-step process, i.e., the formation of a
proto-neutron star and phase conversion.
Neutrino emission in the second step could be significantly
asymmetric since the phase conversion may be off-centered due
to the initial density perturbation (Dong Lai, 2000, personal 
communication). An off-centered transition condition may be 
also realized in the presence of an electron-neutrino-degenerate 
gas in a proto-neutron star (Benvenuto \& Lugones 1999).
Thus the phase transition process may give an additional kick 
to achieve a higher velocity. More detailed investigations are 
desirable to verify these proposals.

We now describe the model in more detail. We assume that the
progenitor of a strange star is surrounded by a huge spherical
comet cloud which is similar to the Oort Cloud in the solar
system. They may be formed during the formation of the massive
star, and have almost finished gravitational relaxation.
Since the progenitor of a strange star should have a mass 
larger than $10M_\odot$, we expect that the radius of the Oort 
Cloud in the progenitor system may be one order of magnitude 
larger than the solar value ($\sim 2\times 10^{13}$ km Weissman 
1990)., i.e., $r_{\rm o}\sim 2\times 10^{14}$ km. 
Supernova explosion blast waves will not
destroy these comet clouds (Tremaine \& Zytkow 1986). The
luminous UV/optical emission from the progenitor is also unlikely
to evaporate the comets. Although the radiation flux received by
the Oort Cloud comets of the massive star should be about a
factor of 30 higher than that received by the Solar Oort Cloud
comets, the existence of copious ``Kuiper Belt'' comets in the
solar system (which is 4 orders of magnitude closer to the sun
than the Oort Cloud) hints that comets can withstand shining with
much higher luminosities. The influence of nearby stars may be 
also not prominent due to the same reason, even if SGRs are
associated with luminous star clusters (e.g. Vrba et al. 2000).
Using the typical proper motion velocity of the SGRs, $V_{_{\rm
SGR}} \sim 10^3 {\rm km~s^{-1}}$, and the typical supernova remnant 
age, $t_{_{\rm SGR}} \sim 10^4$ yr, the distance that a SGR has
traveled since its birth is $r\sim 3\times 10^{14}$ km, remarkably
consistent with the distance of the Oort Cloud $r_{\rm o}$. Thus
the age clustering of the SGRs near $10^4$ yr is simply due to
that this is the age when a lot of impacts are available. The
lack of bursts from the AXPs may be due to their much smaller proper
motion velocities, and probably also their different nature, i.e.,
neutron stars. Although SGR 1806-20 has a smaller
projected proper motion velocity ($V_\perp \sim 100 {\rm
km~s^{-1}}$), we assume that it has a similar velocity as other
SGRs, with a large velocity component along the direction of the
line-of-sight. The capturing rate could be estimated as $\dot N
\sim \pi (2 G M_*/V_{_{\rm SGR}}^2)^2 V_{_{\rm SGR}} n_{\rm c}$,
where $n_{\rm c}$ is the number density of the comets within the
Oort Cloud. To produce a bursting rate of 1 yr$^{-1}$, $n_{\rm
c}$ is required to be $\sim (10^{-22}-10^{-23}) {\rm km}^{-3}$.
This is about 4 orders of magnitude higher than the inferred 
comet number density in the solar Oort Cloud [$\sim (10^{-26}
-10^{-27}) {\rm km}^{-3}$, Weissman 1990], but about 3-4 orders 
of magnitude lower than the inferred number density in the 
Kuiper Belt of solar system
[$\sim (10^{-18}-10^{-20}) {\rm km}^{-3}$, Weissman 1990].
Keeping in mind that the mass density of the Oort Cloud and the
number density of the comets may be enhanced due to accretion
from the dense environment in the supernova remnants (Marsden et
al. 2000) and that the number density quoted for the solar system
might be a lower limit (Weissman 1990), the required $n_{\rm c}$
may be not unreasonable. Some SGRs have more frequent bursting
rate. This may be due to that the strange star has captured a
denser small-scale comet cloud.

When the strange star passes through its Oort Cloud, it may capture
some small-scale clouds and make them circulate around it within
its rest frame\footnote{The fossil disk around the star may be a
good perturber of the comets, which enhances the chances of
captures.}, and the comets within the cloud will be
occasionally accreted onto the strange star surface. The different
bursting luminosities (or more precisely the different energies for
different bursts) correspond to different masses of the impacting
objects. During each impact, the energy released is a sum of the
gravitation energy and the phase conversion energy. The
former has an efficiency of $\eta_{\rm grav}=GM/(R c^2)$, which is
$\sim 20.6\%$ for typical strange star parameters, and the latter
has an efficiency of $\eta_{\rm conv}=\Delta\epsilon /(930 {\rm MeV})$,
where $\Delta\epsilon$ is the energy per baryon released during the
phase conversion. The value of $\Delta\epsilon$ is rather uncertain
which depends on unknown QCD parameters (e.g. MIT bag constant, strange
quark mass and the coupling constant for strong interaction). Some
recent calculations (e.g. Bombaci \& Datta 2000) show that $\Delta
\epsilon \sim 100$ MeV may be reasonable, and we will adopt this 
value for indicative purpose. The deviation of this value from the 
exact value
is not important since this only reflects slightly different required
comet masses. We thus get $\eta_{\rm conv} \sim 11\%$. Assuming that
about one half of the energy will be brought away by neutrinos, the
total $\gamma$-ray emission efficiency is $\eta_\gamma \sim
(\eta_{\rm grav}+\eta_{\rm conv})/2 \sim 16\%$. Thus the repeating
bursts with $L_b \sim 10^{38}-10^{42}{\rm erg~s^{-1}}$ and typical
bursting time $\sim 0.1$ s correspond to the comet masses within the
range of $7\times (10^{16}-10^{20})$ g. These are reasonable values
for comet masses. The so-called giant flare requires an object (an
asteroid or a comet) with a mass of several $10^{24}$ g. In view that
the giant flares are rather rare, it is reasonable to suppose that
such large objects may exist in some dense clouds. Notice
that all the luminosities quoted above are derived under the 
assumption of isotropic emission. For impacting events discussed 
here, during which the emission is anisotropic, the 
required comet masses may be lowered by a factor of 10-100. 
There is no mass distribution data
available for the solar comets, but we expect that the distribution
should be a power law (cf. Pineault \& Poisson 1989). This is
because the stars, which also belong to a gravitationally
self-organized system but in a larger scale, have a well-known
Salpeter's power law mass distribution\footnote{Observationally
the giant flares belong to the high
end of the power-law fluence distributions.}. The bursting intervals
depend on the spatial distribution of the comets within their orbits,
thus there should be no correlations between the luminosity of a
burst and the waiting time before or after this burst. All these
are in excellent agreement with the statistics of the SGR bursts
(G\"ug\"us et al. 1999, 2000). Adopting the typical comet mass as 
the lowest value of the power-law distribution, the comet number 
density inferred above gives a total comet mass of about $0.1 M_\odot$,
not unreasonable due to the same reasons discussed before.

When a comet falls into the strange
star magnetosphere, it will endure tidal distortion and compression
so that they are elongated dense solid objects when they reach the
strange star surface (Colgate \& Petschek 1981). Because they are
globally neutral solid bodies, these comets will not be channeled
to the polar cap regions where the asymptotic accretion flow from
the fossil-disk takes place. This ensures the super-Eddington
luminosity emission from a bare strange star.
The large Coulomb barrier above the bare quark surface (Alcock 
et al. 1986a) will not prevent the object from penetrating
into the quark core. The rising rate of
the energy released from the falling object is similar to the
rising rate of the density from vacuum to solid iron (Howard,
Wilson \& Barton 1981; Katz, Toole \& Unrul 1994), so that the
rising time of the bursts could be of sub-millisecond to
millisecond order,  consistent with the observations of
the giant flares (Hurley et al. 1999). The duration of the hard
spike observed in the giant flares corresponds to the continue
infall time of the object, which is of the order of 0.1-1 s (e.g.
Katz et al. 1994). The August 27 giant flare from SGR 1900+14 has
slightly smaller total energy but both longer rising time and
longer duration of the initial spike than the March 5 event of
SGR 0526-66. This may be understood by assuming that the falling
object of the March 5 event is an asteroid while that of the
August 27 event is a comet, both with a similar mass. During an
impact, both gravitational energy and phase transition energy
will be released in a sufficiently short period of time. Since
there is no baryon contamination for a bare strange star, the
energy will be mainly released as photons and neutrinos. Soon an
optically thick pair fireball will form via the processes such as
$\gamma-\gamma$ (Thompson \& Duncan 1995) and $\gamma-E$ (Usov
1998) processes near a bare quark surface. The magnetic field
will confine this pair plasma, and the soft fading tail of the
giant flares can be due to contraction of this pair bubble
(Thompson \& Duncan 1995; Katz 1996). For the accretion case
discussed here, the energy deposited into the pair bubble is
continually supplied, which is different from the abrupt-release 
case in the magnetar model.
Thus the required magnetic field for confinement is less
demanding, i.e. $B>(2 L_{\rm b} /R^2 c)^{1/2}=8\times 10^{10}
{\rm G}L_{44} ^{1/2}R_6^{-2}$ (Katz 1996). The trapped pair
plasma has a characteristic temperature of $T\sim 23 {\rm keV}$,
and the emergent spectrum is roughly a blackbody with absorption,
which is almost independent on the size of the impacting object
(Katz 1996). All these match the SGR phenomenology well.

Sometimes the accreting matter is not solid, but is an ionized
plasma. In such cases, the effect of the large Coulomb barrier
should be carefully investigated. The kinetic energy of a proton
when it is accreted onto the strange star surface is $E_{\rm k}
\sim G m_{\rm p} M/2R \sim 100$ MeV. However, when materials are
accreted as fluid, it is possible that the kinetic energy will be
radiated away via heat before hitting the surface. In the
accretion column, the scale of the shock wave zone is dependent
on the accretion rate. It is found that when the accretion
luminosity is less than $\sim 4 \times 10^{36} {\rm erg~s^{-1}}$,
the deceleration of the accreting fluid can be neglected (Basko
\& Sunyaev 1976). This is true for SGRs and AXPs since the
quiescent X-ray luminosities of these objects are only
$10^{35}-10^{36} {\rm erg~s^{-1}}$. The Coulomb barrier of a bare
strange star is $E_{\rm C}=(3/4)V_{\rm q} \sim 15$ MeV, where
$V_{\rm q}^3/3\pi^2\sim 20$ MeV is defined as the quark charge
density inside the quark matter (Alcock et al. 1986a). Thus the
accreting fluid, including that from the fossil-disk, can also
penetrate into the strange quark core. This ensures the bare
strange star picture conjectured in this paper. The accreted
matter at the polar cap will undergo phase transition and release
some extra energy. It is unclear whether the slightly harder
spectra of the quiescent emission of the SGRs with respect to the
AXPs is caused by phase transition (we suppose AXPs to be neutron
stars). The enigmatic precursor of the August 29 event of SGR
1900+14 (Ibrahim et al. 2000) may be due to infall of an extended
ionized cloud which is followed by a solid object.

Depending on the impacting angles during the
captures, the small-scale comet clouds may have various orbital
periods and eccentricities, so that the precipitation onto the
star surface is expected to be periodic, especially when the
comets are clustered into a clump in the orbit rather than being
spread over the orbit. In fact, SGR 0526-66 has been reported to
have a 164-day period in bursts (Rothschild \& Lingenfelter
1984). Its present quiescence may be because the previous
comet cloud has been depleted due to many cycles of precipitations.
If it
becomes active again, a different period is expected since it may
have captured a different cloud. SGR 1900+14, on the other hand,
has experienced three active periods during 1979 (Mazets et al.
1981), Jun. - Aug. 1992 (Kouveliotou et al. 1993), and May 1998 -
Jan. 1999 (G\"og\"us et al. 1999). The active periods are short,
and the interval between the first two is roughly twice of that
between the last two (about 6 year). This makes us to suspect a
6-year period for SGR 1900+14 activity. According to this
picture, there should be some bursts in 1986. But this is within
the ``detection gap'' of the SGR bursts when there is no
gamma-ray mission in space before BATSE was launched. {\em Thus
we expect that SGR 1900+14 should become active again during
2004-2005}. This will give a definite test to our model. SGR
1806-20 activity does not have a clear periodicity. However, a
plausible 733-day period is found from its timing residual (Woods
et al. 2000). This might be due to that comets are almost spread
over the whole orbit and the spin-down of the strange star is
perturbed by this comet orbit.

In our model, a fossil-disk is assumed to interpret the spin-down
behavior and the quiescent emission. It is expected that emission
(especially during the bursts) should have some interactions with
the disk with certain optimal geometric configurations. The chance
to see such interactions should be small due to the small size of
the disk. The 6.4 keV emission line from the August 19 burst of
SGR 1900+14 (Strohmayer \& Ibrahim 2000) may be due to the disk's
re-processing the bursting emission.

\section{Discussions}

In this Letter, we propose that the peculiar behaviors of the SGRs
are due to both their Nature (bare strange stars) and their Nurture
(the Oort Cloud in the dense environment). Instead
of invoking the magnetar hypothesis, we adopt the strange star
hypothesis to interpret some interesting features of the SGRs. It
is worth pointing out that the periodic activity does not depend on
the nature of the central star. Although some authors argue that the
bursting phenomenology (e.g. super-Eddington luminosity) can be also
interpreted by colliding comets with a neutron star (e.g. Katz 1996),
we think that a bare strange star is a cleaner interpretation due to
the reasons discussed above. An important criterion to differentiate
our model from the magnetar model is the activity period. If SGR
1900+14 will be turned on again in 2004, the magnetar model is
then not favored, since it may be hard to find a mechanism to
trigger the magnetic field decay instabilities periodically. If it
turns out that some problems (e.g. super-Eddington luminosity,
baryonic contamination, and large proper motion velocity) are not
solvable within the neutron star impacting model, the bursts from
SGR 1900+14 in time then present a support to the strange star
hypothesis and will bring profound implications for fundamental
physics.

According to this picture, there might be some other bare strange
stars which may also have super-Eddington bursts when they collide
with comet-like objects. However, they must have passed through the
Oort Cloud and/or in a much less dense environment, so that the
chance to detect repeating bursts is rare. Single bursting events
are possible and they may account for a small portion of the short,
soft bursts in BASTE data. The association of SGR 1806-20 with a
radio plerion may not be compatible with the present picture, but
recent results indicate that the non-thermal radio core of the
supernova remnant G10.0-0.3 may be associated with another luminous
blue variable rather than with the SGR (Hurley 1999).

\acknowledgments

BZ acknowledges stimulated discussions with Matthew Baring, Zigao 
Dai, Alice Harding, Dong Lai, David Marsden and Vladimir Usov, 
and informative email contacts with Pinaki Chatterjee, Chryssa 
Kouveliotou, Tan Lu and Chris Thompson. RXX thanks
supports from NNSF of China (19803001). GJQ acknowledges supports 
from NNSF of China, the Climbing Project of China, and the Research 
Fund for the Doctoral Program Higher Education. 


\end{document}